\documentclass{elsart}

\usepackage{amsmath,amssymb}

\usepackage[dvips]{graphicx}

\date{}

\allowdisplaybreaks

\begin{document}

\begin{frontmatter}
\title{The critical equation of state of the three-dimensional $O(N)$ universality class: $N>4$.}
\author{Agostino Butti}
\address{Dipartimento di Fisica dell'Universit\`a di Milano-Bicocca \and INFN,
  \mbox{Piazza delle Scienze 3},
  I-20126 Milano,
  Italy
}
\ead{agostino.butti@mib.infn.it}

\author{Francesco Parisen Toldin}
\address{Scuola Normale Superiore \and INFN,
  Piazza dei Cavalieri 7,
  I-56126 Pisa,
  Italy
}
\ead{f.parisentoldin@sns.it}

\begin{abstract}
We determine the scaling equation of state of the three-dimensional
$O(N)$ universality class, for $N=5$, $6$, $32$, $64$. The $N=5$ model
is relevant for the $SO(5)$ theory of high-$T_c$ superconductivity,
while the $N=6$ model is relevant for the chiral phase transition in
two-color QCD with two flavors.
We first obtain the critical exponents and the small-field, high-temperature, expansion of the effective potential (Helmholtz free energy) by analyzing the available perturbative series, in both fixed-dimension and $\epsilon$-expansion schemes. Then, we determine the critical equation of state by using a systematic approximation scheme, based on polynomial representations valid in the whole critical region, which satisfy the known analytical properties of the equation of state, take into account the Goldstone singularities at the coexistence curve and match the small-field, high-temperature, expansion of the effective potential. This allows us also to determine several universal amplitude ratios. We also compare our approximate solutions with those obtained in the large-$N$ expansion, up to order $1/N$, finding good agreement for $N\gtrsim 32$.
\end{abstract}


\end{frontmatter}

\section{Introduction}
According to the Renormalization Group theory, it is possible to
classify continuous phase transitions into universality classes,
determined by few global properties, such as space dimensionality,
number of components of the order parameter, the range of interaction,
symmetry group and symmetry-breaking pattern \cite{ZJbook,phase14book}. Some physical quantities, called universal, are independent from the microscopic structure of the interaction: models belonging to the same universality class have the same critical exponents, scaling functions and universal amplitude ratios. One of the most important universality classes are the $O(N)$ classes, which are characterized by a $N$-component order parameter, and a symmetry group $O(N)$ which is spontaneously broken to a subgroup $O(N-1)$ in the low-temperature phase. For a recent review on this subject, see Ref. \cite{review}. This universality class can describe the critical behavior of many physical systems, which undergo a second order phase transition. For instance, the liquid-vapor transition in simple fluids and the magnetic transition in uniaxial (anti-)ferromagnets belong to the Ising universality class ($N=1$). The XY model, which corresponds to $N=2$, describes the helium superfluid transition and the Meissner transition in type-II superconductors. If the order parameter has $N=3$ components, we get the Heisenberg universality class, which describes the critical behavior of isotropic magnets. The three-dimensional $O(4)$ model is related to the finite-temperature phase transition in QCD with two light flavors \cite{PW-84,Wilczek-92,RW-93,Butti:2003nu}.

In this article, we study the critical equation of state and universal quantities for the $O(N)$ three-dimensional universality class, with $N > 4$. We first consider the $N=5$ and the $N=6$ models. The $3$-$D$ $O(5)$ model is relevant for the so-called $SO(5)$ theory of high-$T_c$ superconductivity \cite{SO5}, since, according to this theory, the $SO(5)$ symmetry should be approximately realized at a multicritical point of the phase diagram\footnote{However, it has been shown \cite{Calabrese:2002bm} that the $O(5)$ multicritical fixed point is in fact unstable: this implies that the $SO(5)$ symmetry is not realized asymptotically at the multicritical point, but it controls the critical behavior in a preasymptotic region for $t \approx 10^{-2}$.}. According to universality arguments, the $N=6$ case should describe the chiral phase transition in QCD with two flavors and two colors \cite{Smilga:1994tb}. This subject has been investigated by using an effective field theory in Ref. \cite{Wirstam:1999ds}. For all these $O(N)$ models, we first analyze the perturbative series available, in both fixed-dimension scheme \cite{Baker_series,sokolov_exp,sokolov_r2n} and $\epsilon$-expansion \cite{Gorishnii,Kleinert,Pelissetto:1997gk,Pelissetto:1998sk,potential2000};
by resumming these series we determine the critical exponents and the
coefficients $r_6$ and $r_8$ of the small-magnetization expansion of the
critical equation of state in the high-temperature phase. In
fixed-dimension we resum the series for the critical exponents $\eta(g)$,
$\eta_t(g)$ and for the beta function $\beta(g)$ up to six loops, and the
series for $r_6(g)$ and $r_8(g)$ up to four and three loops respectively.
In $\epsilon$-expansion we resum the series for $\eta(\epsilon)$ and
$\eta_2(\epsilon)$ up to five loops and the series for $r_6(\epsilon)$,
$r_8(\epsilon)$ and for the universal quantity $g_4(\epsilon)$ up to
three loops.
Then, we implement a systematic approximation scheme for the equation of state, which provides an analytic continuation to the low-temperature phase. This procedure, based on polynomial representations, has already been used for other models in the $O(N)$ universality class \cite{ising,XY2000,XY2001,heisenberg,O4}. We determine, using this approximation scheme, the scaling functions and other universal quantities. We also compare the results with those in the large-$N$ limit, up to order $1/N$, that we calculate starting from the large-$N$ expansion of the equation of state reported in Ref. \cite{largeNeq}. A recent review on the subject of large-$N$ field theory can be found in Ref. \cite{Moshe:2003xn}. In order to check the convergence of this expansion, we repeat our work for two other $O(N)$ models, $N=32$ and $N=64$. In these cases we find a good agreement of our results with those obtained in the $1/N$ expansion.

This paper is organized as follows. In section \ref{notation} we summarize the scaling properties of the equation of state for the $O(N)$ model.

In section \ref{amplitudes} we introduce the notation and define some universal amplitude ratios that we calculate.

In section \ref{parametric} we introduce a parametric representation of the equation of state, valid in the whole critical region.

In section \ref{approx_eq} we consider a systematic approximation scheme, based on polynomials, that we use in our calculations.

In section \ref{critical56} we show critical exponents, universal couplings and other various universal quantities for the $O(5)$ and $O(6)$ three-dimensional models. We present a comparison of this results with those available in literature and with the $1/N$ expansion. We also plot the universal scaling functions we have obtained.

In section \ref{criticalNlarge} we consider two models with large $N$: the $N=32$ and the  $N=64$ three-dimensional models. We report various universal quantities for these models, comparing the results with those obtained from the $1/N$ expansion of the equation of state.

In appendix \ref{resummation} we report the algorithm and formulas that we have used for the analysis of the perturbative series.

In appendix \ref{largeN} we report the $1/N$ expansion of the equation of state and the relevant formulas for the universal quantities.

\section{The scaling equation of state}
\label{notation}
The equation of state is a relation between the magnetization $\vec{M}$, the reduced temperature $t \equiv (T-T_c)/T_c$ and the external magnetic field $\vec{H}$. It is obtained by noting that $\vec{H}$ is the derivative of the effective potential (Helmholtz free energy) $A(t,M)$ with respect to $\vec{M}$. Near the critical point $\vec{H}=0$, $T=T_c$, it can be written \cite{ZJbook,phase14book} in the scaling form
\begin{equation}
\label{scaling}
\vec{H} = (B^c)^{-\delta} \vec{M} M^{\delta - 1} f(x), \qquad x \equiv B^{1/\beta}\, t M^{-1/\beta},
\end{equation}
where $M\equiv |\vec{M}|$, $f(x)$ is a universal scaling function, fixed by the normalization $f(0)=1$, $f(-1)=0$, $B^c$ and $B$ are the non-universal magnetization amplitudes at the critical isotherm and at the coexistence curve:
\begin{align}
\label{isotherm_crit}
\vec{M} &= B^c \vec{H} H^{1/\delta-1}, \qquad t=0,\\
\label{coex_crit}
M &= B (-t)^\beta, \qquad t<0, \ H = 0.
\end{align}
In the high-temperature phase $t > 0$, the critical equation of state can also be written as:
\begin{equation}
\label{scalingF}
\vec{H} = ab \frac{\vec{M}}{M} t^{\beta\delta} F(z), \qquad z \equiv b M t^{-\beta},
\end{equation}
where $F(z)$ is a universal scaling function. Since the free energy is analytic in the plane $(t,H)$ outside the coexistence curve and the critical point, $F(z)$ can be expanded around $z=0$ in odd powers. The non-universal constants $a$ and $b$, which appear in (\ref{scalingF}), fix the normalization on $F(z)$ such that
\begin{equation}
\label{expF}
F(z) = z + \frac{z^3}{3!} + \sum_{n \geq 3} \frac{r_{2n}}{(2n-1)!} z^{2n-1}.
\end{equation}
In the following, we refer to the universal coefficients $r_{2n}$ as the universal couplings. Inverting (\ref{scaling}), the equation of state can also be written as
\begin{equation}
\label{scalingE}
\vec{M} = B_c \vec{H} H^{1/\delta-1} E(y), \qquad y \equiv (B/B_c)^{1/\beta}tH^{-1/(\beta\delta)},
\end{equation}
where $E(y)$ is a universal scaling function. The relation between the scaling functions $E(y)$ and $f(x)$ is:
\begin{equation}
E(y) = f(x)^{-1/\delta} , \qquad y = x f(x)^{-1/(\beta + \gamma)}.
\end{equation}
From (\ref{expF}) it follows the large-$x$ behavior of $f(x)$:
\begin{equation}
f(x) = x^\gamma \sum_{n \geq 0} f_n^\infty x^{-2n\beta}.
\end{equation}
The fact that the equation of state is analytical for $H>0$ implies that $f(x)$ can be expanded around $x=0$
\begin{equation}
f(x)=1+\sum_{n=1}^{\infty} f_n^0 x^n,
\end{equation}
and that for large-$z$:
\begin{equation}
F(z) = z^\delta \sum_{n\geq 0}F_n^\infty z^{-n/\beta}.
\end{equation}
The relation between $F(z)$ and $f(x)$ is:
\begin{equation}
z^{-\delta}F(z) = F_0^\infty f(x), \qquad z=z_0 x^{-\beta},
\end{equation}
where
\begin{equation}
z_0 = b B = (R_4^+)^{1/2}
\end{equation}
is a universal amplitude ratio (see section \ref{amplitudes}).

At the coexistence curve, that is for $x\rightarrow -1$, spontaneous
breaking of the $O(N)$ symmetry occurs, so $f(-1)=0$. In three
dimensions, the presence of Goldstone modes leads to the prediction
\cite{Brezin:qa,largeNeq,Lawrie:vk}:
\begin{equation}
\label{fx_coex}
f(x) \simeq c_f (x+1)^2, \qquad x\rightarrow -1.
\end{equation}
This behavior is exact in the large-N limit \cite{largeNeq}. The correction to
(\ref{fx_coex}) is less clear. If we define $v\equiv x+1$ and
$w=HM^{-\delta}$, using $\epsilon$-expansion it has been
conjectured \cite{Lawrie:vk} that $v$ admits a double
expansion in powers of $w$ and $w^{(d-2)/2}$:
\begin{equation}
\label{conj}
v=c_1 w^{(d-2)/2} + b_1 w + c_2 w^{d-2} + b_2 w^2 + \ldots
\end{equation}
In $d=3$ eq. (\ref{conj}) implies that $v$ has an expansion in powers of
$w^{1/2}$. However, this is not true in the large-$N$ limit. In fact the $1/N$
correction to the leading behavior \cite{Pelissetto:1999cq} shows the
presence of logarithms for $x\rightarrow -1$.

From (\ref{fx_coex}) one can derive that the transverse and
longitudinal susceptibilities, near the critical point, for $t<0$ and $H\rightarrow 0$,
diverge as:
\begin{align}
\chi_T &= \frac{M}{H}\\
\chi_L &= \frac{\partial M}{\partial H} \propto (-t)^{\beta(1-\delta(d-2)/2)}H^{(d-4)/2}.
\end{align}
In particular, in three dimensions $\chi_L$ diverges as $H^{-1/2}$ for
$H\rightarrow 0$.
The scaling behavior of the longitudinal susceptibility in the
critical domain can be derived from (\ref{scalingE}):
\begin{equation}
\chi_L = B_c H^{1/\delta - 1} D(y),
\end{equation}
where the scaling function $D(y)$ is:
\begin{equation}
D(y) = \frac{1}{\delta}\left[E(y)-\frac{y}{\beta}E'(y)\right] = \frac{\beta f(x)^{1-1/\delta}}{\beta\delta f(x) - xf'(x)}.
\end{equation}
This function has a maximum at $y=y_{max} > 0$, which defines the
crossover or pseudo-critical line \cite{review} (see section \ref{amplitudes}).

\section{Universal amplitude ratios}
\label{amplitudes}
Aside from scaling functions defined in the previous paragraph, there
are other important universal quantities: the amplitude ratios. They
can be defined from the critical behavior of zero-momentum
quantities, such as specific heat, susceptibility, etc.\ldots

Near the critical point, the specific heat has the following singular
behavior:
\begin{equation}
\label{heat_crit}
T_c C_H=A^\pm|t|^{-\alpha} + b,
\end{equation}
where the amplitudes $A^+$ and $A^-$ are for $T>T_c$ and $T<T_c$,
respectively, and $b$ is a non-universal constant (we use the notations of Ref. \cite{review}). The susceptibility, in the high-temperature phase, diverges for $t\rightarrow 0^+$ as:
\begin{equation}
\label{susc_crit}
\chi=NC^+ t^{-\gamma}.
\end{equation}
The critical behavior of the four-point connected function, at zero
momentum, is given, in the high-temperature phase, by:
\begin{equation}
\label{4point_crit}
\chi_4 =  \frac{N(N+2)}{3}C_4^+ t^{-\gamma-2\beta\delta}.
\end{equation}
The divergence of the correlation length at the critical point is
described by:
\begin{equation}
\label{length_crit}
\xi=f^+ t^{-\nu}.
\end{equation}
Universal amplitudes can also be obtained by considering the
crossover, or pseudo-critical, line \cite{review}, which is defined as the reduced
temperature $t_{max}(H)$ where the longitudinal susceptibility has a
maximum at $H$ fixed:
\begin{align}
\label{crossover_crit}
t_{max}(H)&=T_p H^{1/(\beta+\gamma)},\\
\chi_L(t_{max},H)&=C_p t_{max}^{-\gamma}.
\end{align}
Corresponding to this maximum, the scaling variables $x$, $y$ and $z$ have the universal values $x_{max}$, $y_{max}$, and $z_{max}$, respectively.
From equations (\ref{isotherm_crit}), (\ref{coex_crit}) and (\ref{heat_crit}-\ref{crossover_crit}), one can define several
universal amplitude ratios.
Those involving the correlation length amplitude $f^+$ can be obtained by using the zero-momentum four-point coupling $g_4$, defined as:
\begin{equation}
g_4 \equiv -\frac{3N}{N + 2} \frac{\chi_4}{\chi^2 \xi^d}.
\end{equation}
At the critical point, $g_4$ approaches a universal value, that we indicate with the same symbol.
Several definitions of universal amplitude ratios \cite{review} are reported in table \ref{amplitudes_def}, as well as the universal quantity $g_4$.

\begin{table}
\caption{Definition of universal amplitude ratios and universal quantity $g_4$ \protect{\cite{review}}.}
\footnotesize
\begin{tabular}{|ll|}
\hline
\label{amplitudes_def}
$U_0\equiv A^+/A^-$
&
$R_\alpha \equiv (1-U_0)/\alpha$
\\
$R_c^+\equiv \alpha A^+C^+/B^2$
&
$R_4^+\equiv - C_4^+B^2/(C^+)^3$
\\
$R_\chi\equiv C^+ B^{\delta-1}/B_c^\delta$
&
$R^+_\xi\equiv (\alpha A^+)^{1/3}f^+$
\\
$P_m \equiv { T_p^\beta B/B^c} $
&
$P_c \equiv -T_p^{2\beta\delta}C^+/C_4^+$
\\
$ R_p \equiv { C^+/C_p} $
&
$g_4 \equiv -C_4^+/[(C^+)^2(f^+)^3]$
\\
\hline
\end{tabular}
\end{table}

\section{Parametric representation of the equation of state}
\label{parametric}
In order to study the equation of state in the whole critical regime, it is quite useful to introduce a parametric representation that implements the known analytical and scaling properties \cite{parrep}. We introduce two variables, $R$ and $\theta$, such that:
\begin{equation}
\label{par}
\left\{
\begin{aligned}
M &= m_0 R^\beta m(\theta)\\
t &= R (1-\theta^2)\\
H &= h_0 R^{\beta\delta}h(\theta),
\end{aligned}
\right.
\end{equation}
where $h_0$ and $m_0$ are two normalization constants. $R$ is a nonnegative variable which measures the distance from the critical point. The functions $h(\theta)$ and $m(\theta)$ are odd and are conventionally normalized so that $h(\theta)=\theta + O(\theta^3)$ and $m(\theta)=\theta + O(\theta^3)$, for $\theta \rightarrow 0$. From (\ref{scalingF}), (\ref{expF}) and this normalization, one can obtain that $h_0/m_0=ab^2$; in order to take account of this, we define a parameter $\rho$ such that
\begin{equation}
h_0=\rho ab, \qquad m_0=\rho/b.
\end{equation}
As can be seen from (\ref{par}), the line $\theta=0$ corresponds to the high-temperature phase, while on the critical isotherm $\theta=1$. The coexistence curve is given by $\theta=\theta_0$, the first positive zero of $h(\theta)$. Clearly, consistence of the representation requires that $m(\theta)>0$ for $0 < \theta\leq\theta_0$ and $h(\theta)>0$ for $0 < \theta < \theta_0$. The mapping (\ref{par}) is not invertible when its Jacobian vanishes; this happens when:
\begin{equation}
Y(\theta)\equiv (1-\theta^2)m'(\theta)+2\beta\theta m(\theta) = 0.
\end{equation}
Defining $\theta_l$ as the smallest zero of $Y(\theta)$, it must be $\theta_l>\theta_0$.

All the scaling functions, scaling variables and universal amplitude ratios definitions, that we have reported in sections \ref{notation} and \ref{amplitudes}, can be written in terms of the representation (\ref{par}); the formulas can be found in \cite{heisenberg}. We only note that, in order to reproduce the expected behavior (\ref{fx_coex}) of $f(x)$ near $x=-1$, the zero $\theta_0$ of $h(\theta)$ must be double:
\begin{equation}
h(\theta) \sim (\theta-\theta_0)^2, \qquad \theta\rightarrow\theta_0.
\end{equation}

\section{Approximate equation of state}
\label{approx_eq}
Following \cite{XY2000}, we now introduce an approximate polynomial representation, based on the parametric representation (\ref{par}), that has all the expected properties we enumerated in section \ref{parametric}. It extends the one considered in \cite{ising} and was already used in other three-dimensional $O(N)$ systems: the $XY$ model \cite{XY2000,XY2001} ($N=2$), the Heisenberg model \cite{heisenberg} ($N=3$) and the $O(4)$ model \cite{O4}.
In this approximation scheme\footnote{In \cite{XY2000,XY2001,heisenberg,O4}  it was also considered another approximation scheme, denoted by $(B)$. However, in this work, the scheme $(B)$ didn't work, either by providing $\theta_l < \theta_0$ or by failing to satisfy eq. (\ref{consistency}). We mention that also in the $N=4$ model, using the approximation schemes $(B)$ and that of eq. (\ref{schemeAm}), the latter provided better results \cite{O4}.}, we take $h(\theta)$ as a 5th order polynomial and $m(\theta)$ as a polynomial of degree $2n+1$:
\begin{equation}
\label{schemeAm}
\begin{split}
m(\theta) &= \theta \left( 1+ \sum_{i=1}^n c_i\theta^{2i}\right),\\
h(\theta) &= \theta \left( 1- \frac{\theta^2}{\theta_0^2} \right)^2.
\end{split}
\end{equation}
It can be shown that, in the \mbox{$N\rightarrow\infty$} limit, the scheme $n=0$ is exact \cite{largeNeq}. The parameters $\rho$, $\theta_0$ and $n$ coefficients $c_i$, i.e. $n+2$ parameters, are fixed by imposing that the equation of state, in the limit $z\rightarrow 0$, or, equivalently, $\theta\rightarrow 0$, reproduces the expansion (\ref{expF}). It is easy to see that we can implement the scheme at the order $n$ if we know $n+1$ coefficients $r_{2k}$: this occurs because imposing the small-$z$ expansion (\ref{expF}) at the first order gives a trivial equation.

We see that there are two sources of error in this calculus. The first one is the uncertainty on the input parameters. The second one is the systematic error due to this approximation scheme. Indeed, if we are able to compute at least the $n=0$ and $n=1$ schemes, we can estimate this error by looking at the differences between the results for $n=0$ and those for $n=1$. Taking into account equations (\ref{schemeAm}), we see that consistence of the whole calculus requires the coefficients $c_i$ to be small. In particular, for the $n=1$ case, it must be:
\begin{equation}
\label{consistency}
|c_1| \theta_0^2 \ll 1.
\end{equation}

\section{Critical equation of state for the $O(5)$ and $O(6)$ models}
\label{critical56}
We present in table \ref{exp56} the results for critical exponents and universal couplings for the models $N=5$ and $N=6$, obtained by an analysis of the fixed-dimension perturbative series \cite{Baker_series,sokolov_exp,sokolov_r2n} and of the $\epsilon$-expansion series \cite{Gorishnii,Kleinert,Pelissetto:1997gk,Pelissetto:1998sk,potential2000}.
In fixed-dimension we resum the series for the critical exponents $\eta(g)$,
$\eta_t(g)$ and for the beta function $\beta(g)$ up to six loops, and the
series for $r_6(g)$ and $r_8(g)$ up to four and three loops respectively.
In $\epsilon$-expansion we resum the series for $\eta(\epsilon)$ and
$\eta_2(\epsilon)$ up to five loops and the series for $r_6(\epsilon)$,
$r_8(\epsilon)$ and $g_4(\epsilon)$ up to three loops.
We also show a comparison with other determinations available in literature and with the large-$N$ expansion. The series for $r_8$, in the fixed-dimension scheme, is known up to three loops \cite{sokolov_r2n}: for this reason we can only provide a rough estimate of this coupling. The resummation of universal couplings in \mbox{$\epsilon$-expansion} is constrained with the exact results available in dimensions $d=0$ and $d=1$. The resummation is performed with the conformal mapping method; see appendix \ref{resummation} for details.

\begin{table*}
\caption{Critical exponents, universal couplings and the universal quantity $g_4$ for the models $N=5$ and $N=6$ obtained by resummation of the fixed-dimension series \protect{\cite{Baker_series,sokolov_exp,sokolov_r2n}} (first row) and $\epsilon$-expansion series \protect{\cite{Gorishnii,Kleinert,Pelissetto:1997gk,Pelissetto:1998sk,potential2000}} (second row). A comparison with other determinations is shown.}
\footnotesize
\begin{tabular}{|cl|c|c|c|c|c|c|}
\hline
\label{exp56}
$N$ &                & $g_4$      & $\eta$     & $\nu$       & $r_6$     & $r_8$ \\
\hline
$5$ & fixed-dim      & $15.74(2)$ & $0.030(1)$ & $0.764(2)$  & $1.72(2)$ & $-1(3)$ \\
    & $\epsilon$-exp & $15.6(1)$  & $0.034(2)$ & $0.764(6)$ & $1.70(1)$ & $-0.3(5)$ \\
    &                & $^a 15.69$ & $^a 0.034$ & $^a 0.766$  & $^b 1.73$ & \\
    &                &            & $^k 0.054$ & $^k 0.784$  & $^k 2.88$ & $^k -13$\\
\hline
$6$ & fixed-dim      & $14.43(4)$            & $0.028(1)$ & $0.788(2)$  & $1.66(2)$ & $-1(3)$ \\
    & $\epsilon$-exp & $14.23(8)$            & $0.032(2)$ & $0.790(6)$ & $1.64(2)$ & $-0.7(4)$ \\
    &                & $^a 14.39$            & $^a 0.031$ & $^a 0.790$ & $^b 1.67$ & $^b -0.28$ \\
    &                & $^c 14.59^{+5}_{-11}$ & $^e 0.032(10)$ & $^e 0.804(3)$ &&\\
    &                & $^d 14.65^{+5}_{-11}$ & $^f 0.034(9)$ & $^f 0.821(3)$ &&\\
    &                &                       & $^g 0.033(9)$ & $^g 0.796(3)$ &&\\
    &                &                       & $^h 0.037(9)$ & $^h 0.819(3)$ &&\\
    &                &                       & $^i 0.031(5)$ & $^i 0.786(5)$ &&\\
    &                &                       & $^j 0.039(3)$ & $^j 0.818(5)$ &&\\
    &                &                       & $^k 0.045$ & $^k 0.820$ & $^k 2.54$ & $^k -11$\\
\hline
\end{tabular}\\
$^a$\cite{sokolov_exp} FT $d=3$ exp,
$^b$\cite{sokolov_r2n} FT $d=3$ exp\\
$^c$\cite{ButeraComi_couplings} HT sc,
$^d$\cite{ButeraComi_couplings} HT bcc\\
$^e$\cite{ButeraComi_exponents} HTE sc unbiased,
$^f$\cite{ButeraComi_exponents} HTE sc $\theta$-biased\\
$^g$\cite{ButeraComi_exponents} HTE bcc unbiased,
$^h$\cite{ButeraComi_exponents} HTE bcc $\theta$-biased\\
$^i$\cite{Loison} MC,
$^j$\cite{MCO6} MC\\
$^k$Large-$N$ expansion to order $O(1/N)$, see appendix \ref{largeN}\\
\end{table*}

\begin{figure}
\begin{minipage}{0.5\linewidth}
\includegraphics[keepaspectratio, width=0.93\linewidth]{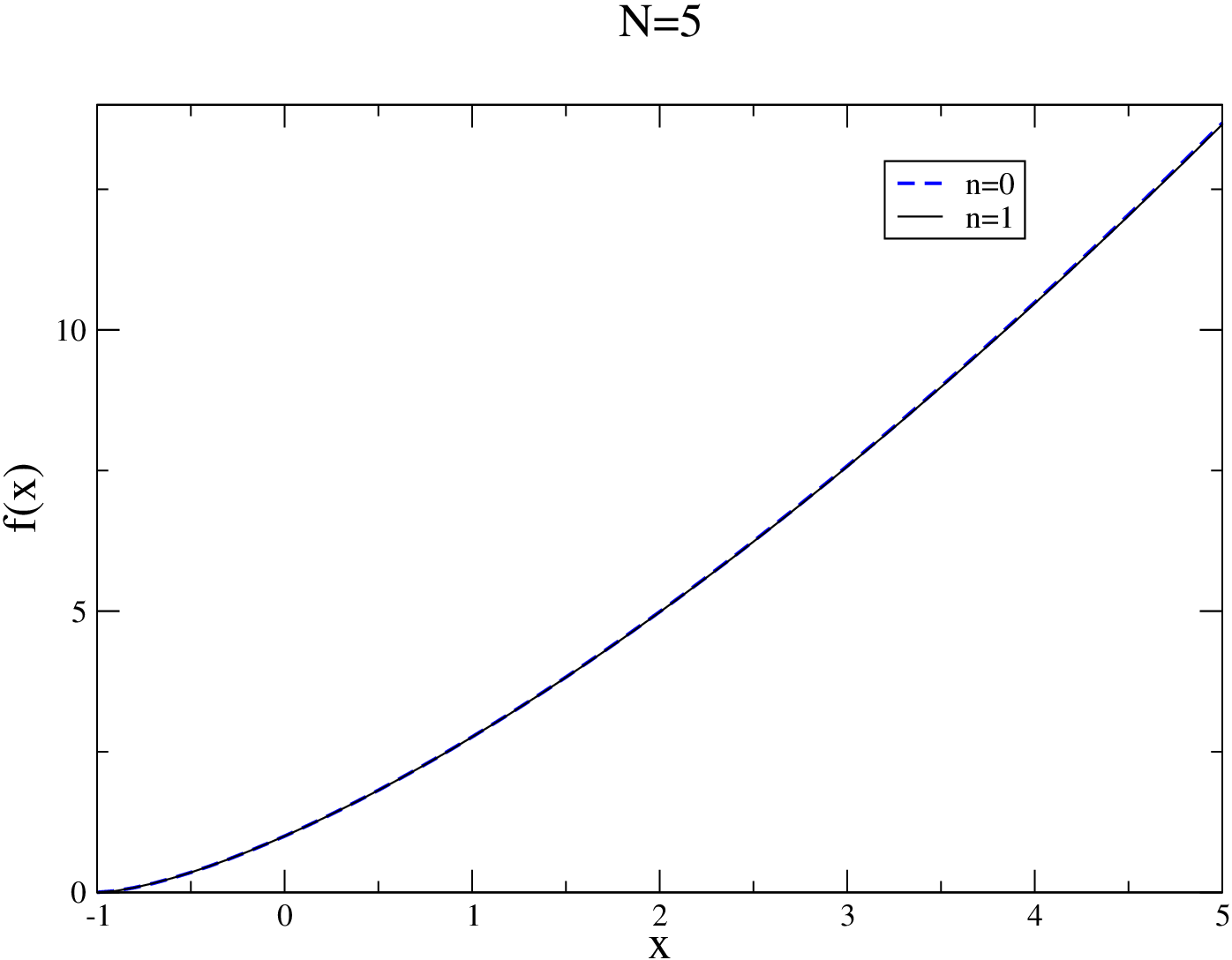}
\end{minipage}
\begin{minipage}{0.5\linewidth}
\includegraphics[keepaspectratio, width=0.93\linewidth]{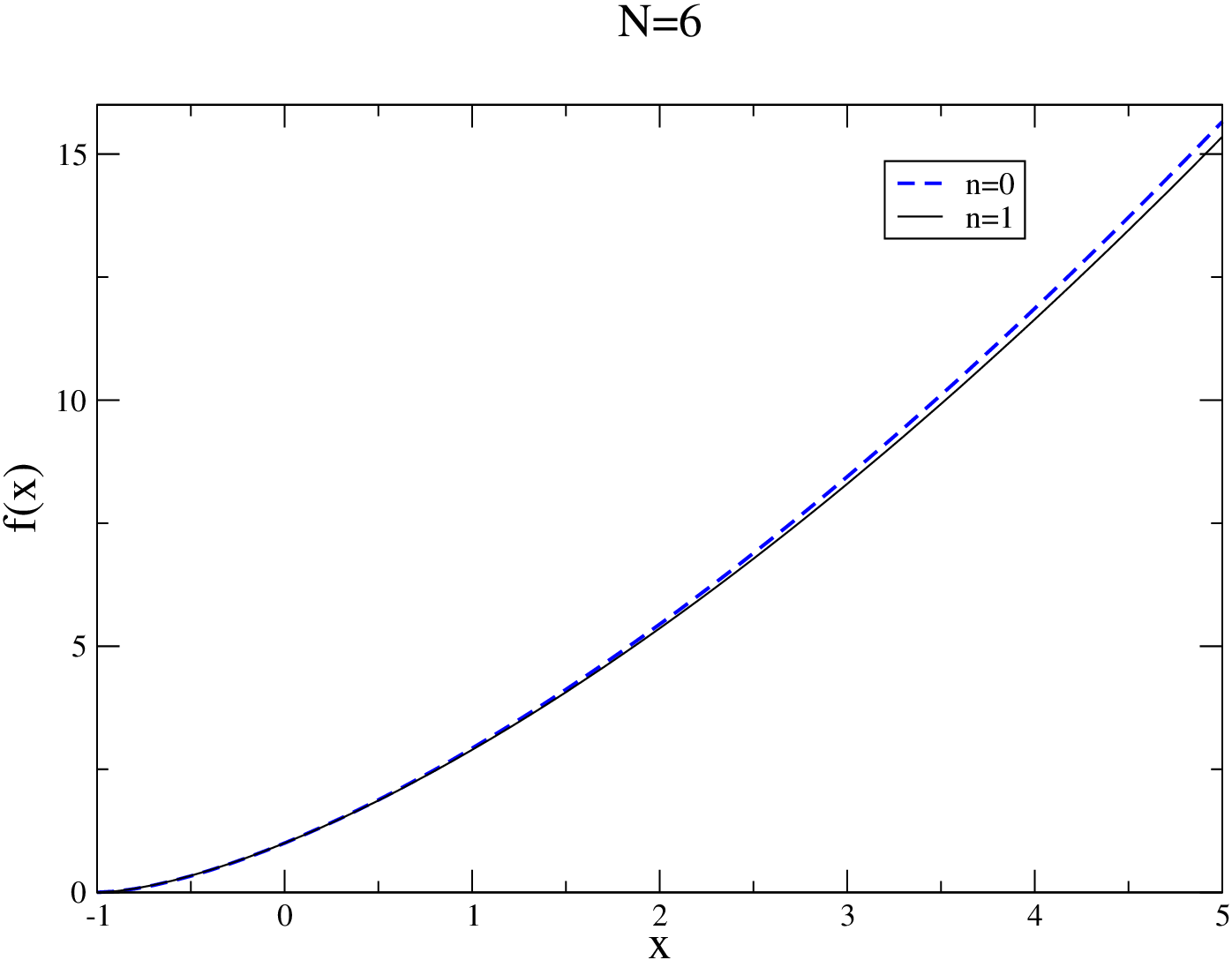}
\end{minipage}
\\
\caption{The scaling function $f(x)$ for the models $N=5$ and $N=6$.}
\label{figfx}
\end{figure}

\begin{figure}
\begin{minipage}{0.5\linewidth}
\includegraphics[keepaspectratio, width=0.93\linewidth]{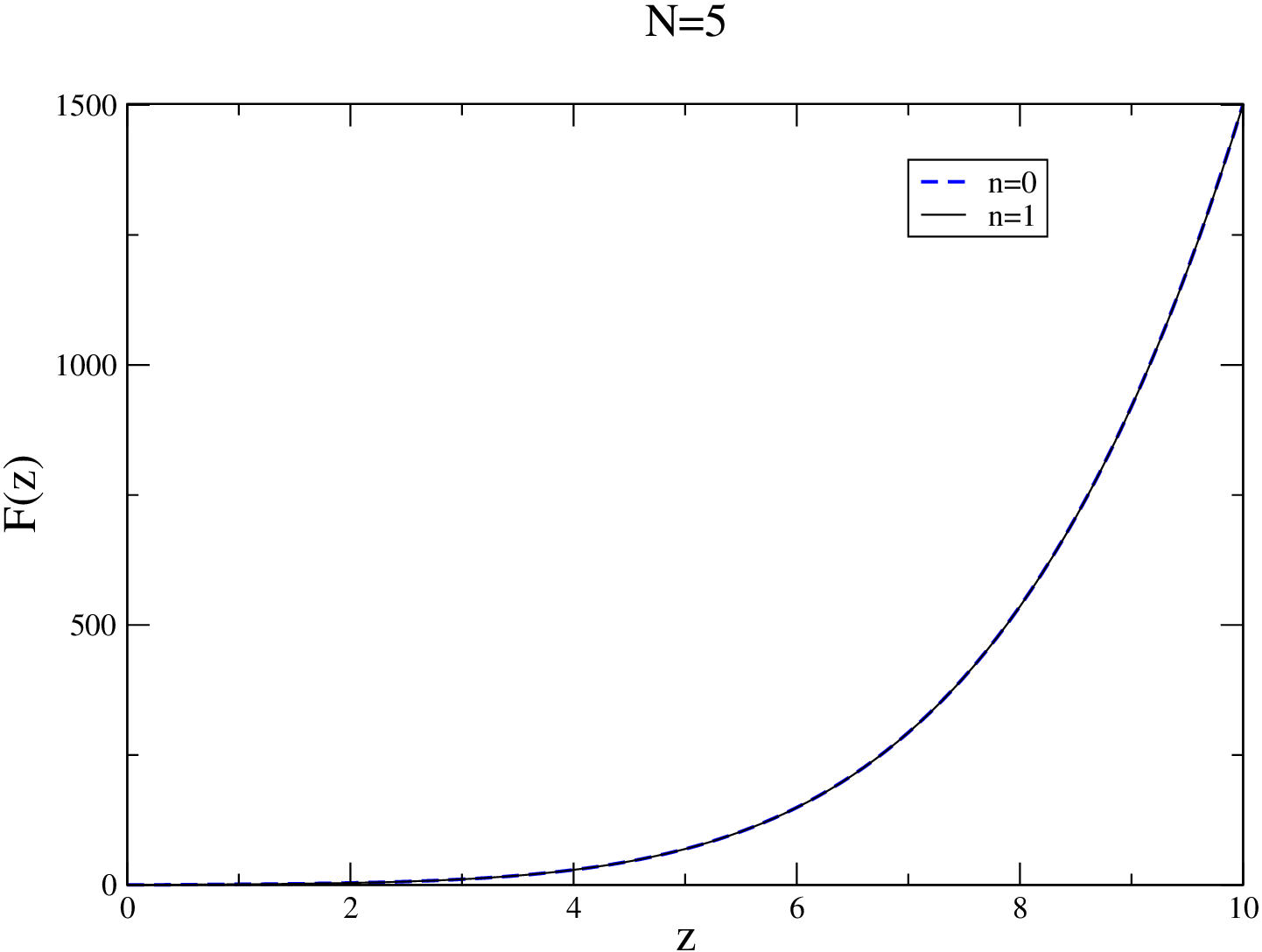}
\end{minipage}
\begin{minipage}{0.5\linewidth}
\includegraphics[keepaspectratio, width=0.93\linewidth]{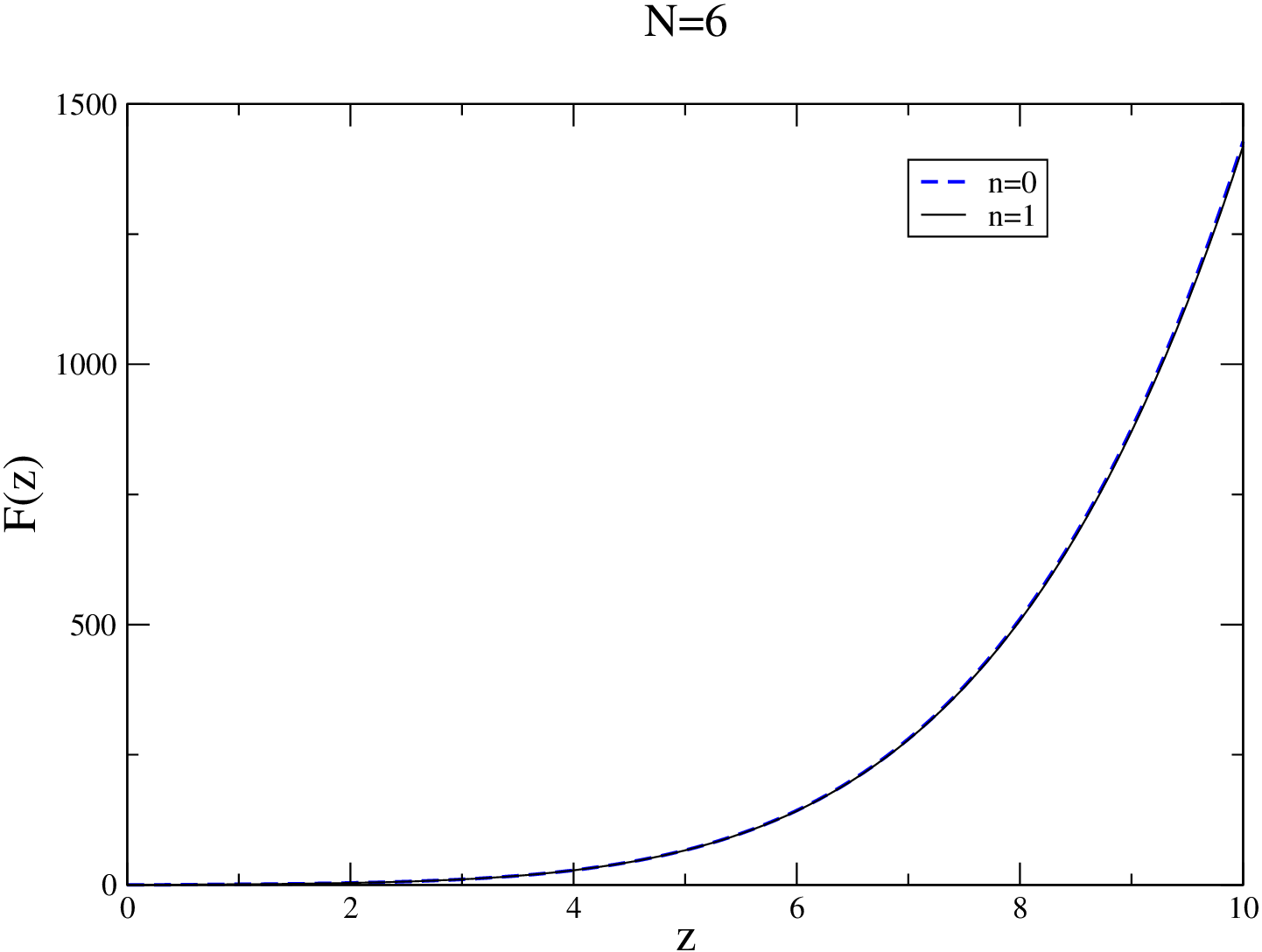}
\end{minipage}
\\
\caption{The scaling function $F(z)$ for the models $N=5$ and $N=6$.}
\label{figFz}
\end{figure}

\begin{figure}
\begin{minipage}{0.5\linewidth}
\includegraphics[keepaspectratio, width=0.93\linewidth]{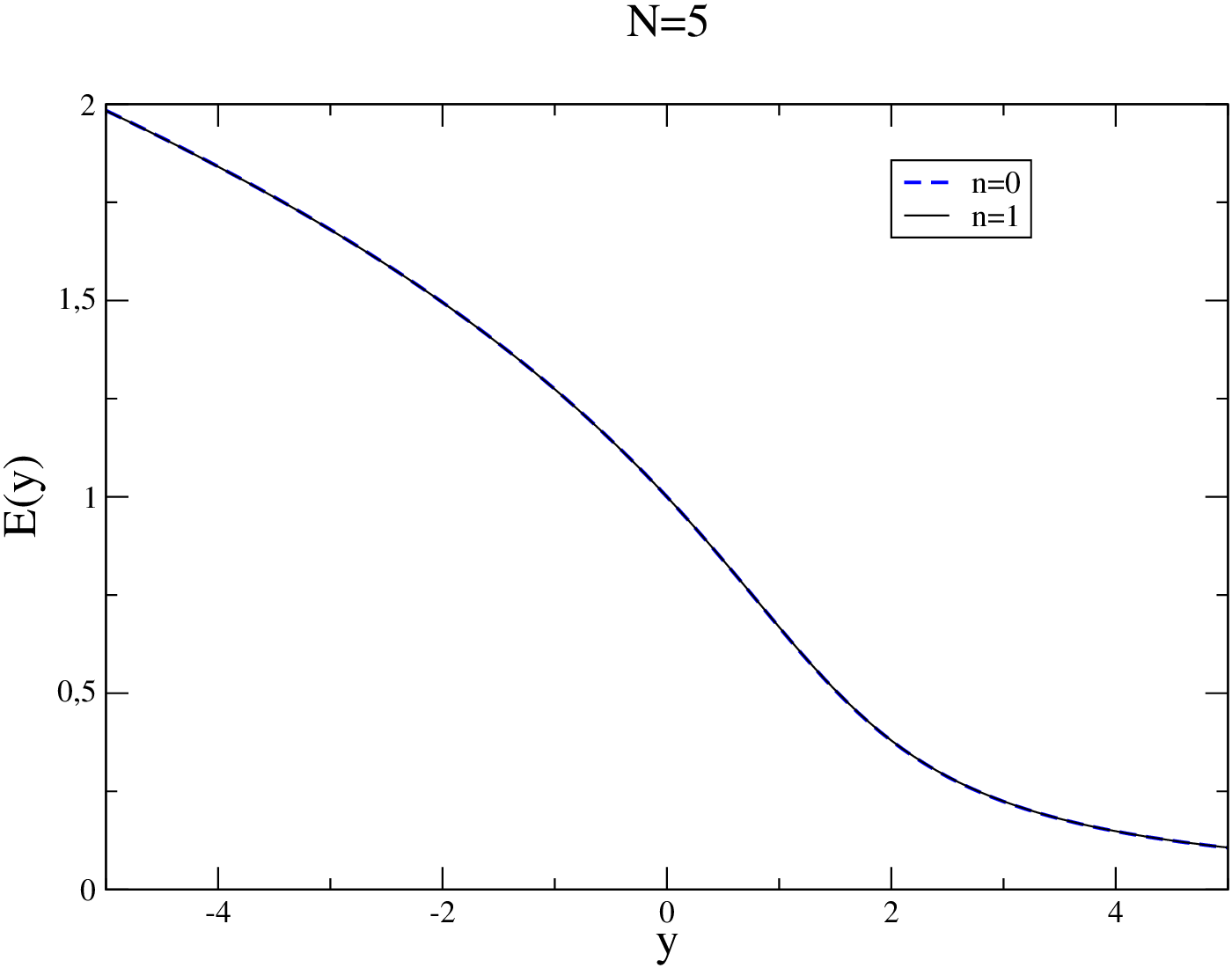}
\end{minipage}
\begin{minipage}{0.5\linewidth}
\includegraphics[keepaspectratio, width=0.93\linewidth]{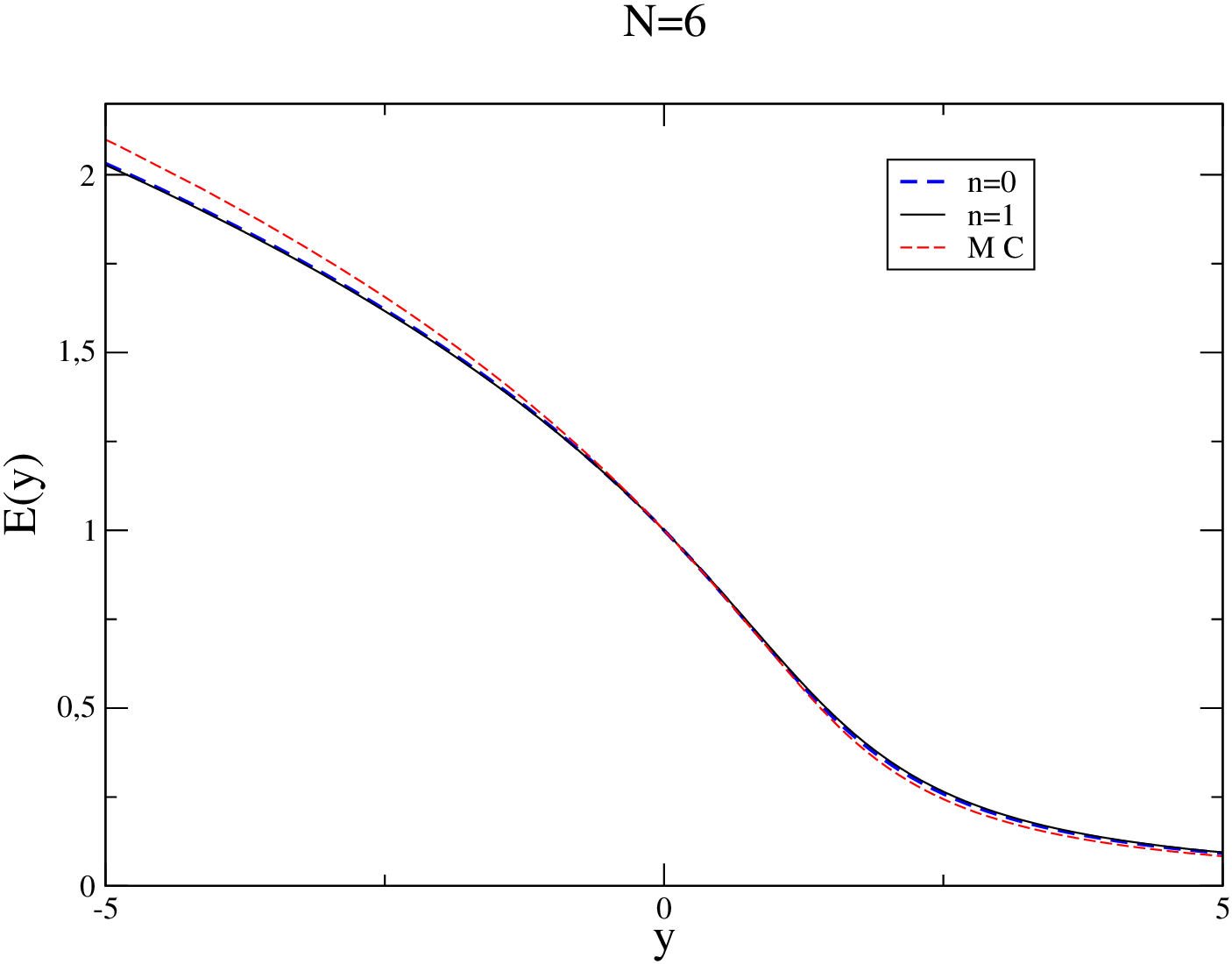}
\end{minipage}
\\
\caption{The scaling function $E(y)$ for the models $N=5$ and $N=6$. For the $O(6)$ model we show, for comparison, the results of Ref. \protect{\cite{MCO6}}.}
\label{figEy}
\end{figure}

\begin{figure}
\begin{minipage}{0.5\linewidth}
\includegraphics[keepaspectratio, width=0.93\linewidth]{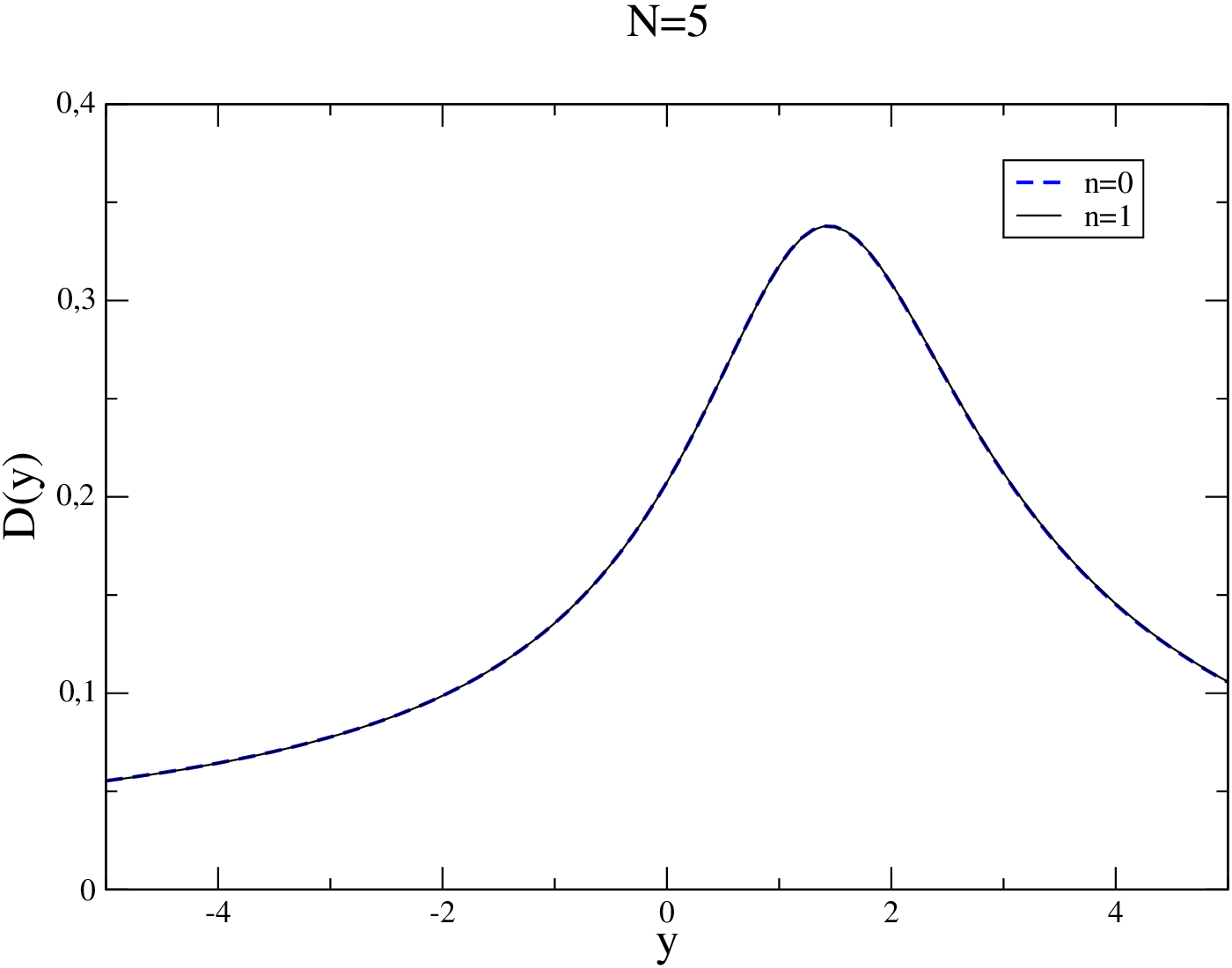}
\end{minipage}
\begin{minipage}{0.5\linewidth}
\includegraphics[keepaspectratio, width=0.93\linewidth]{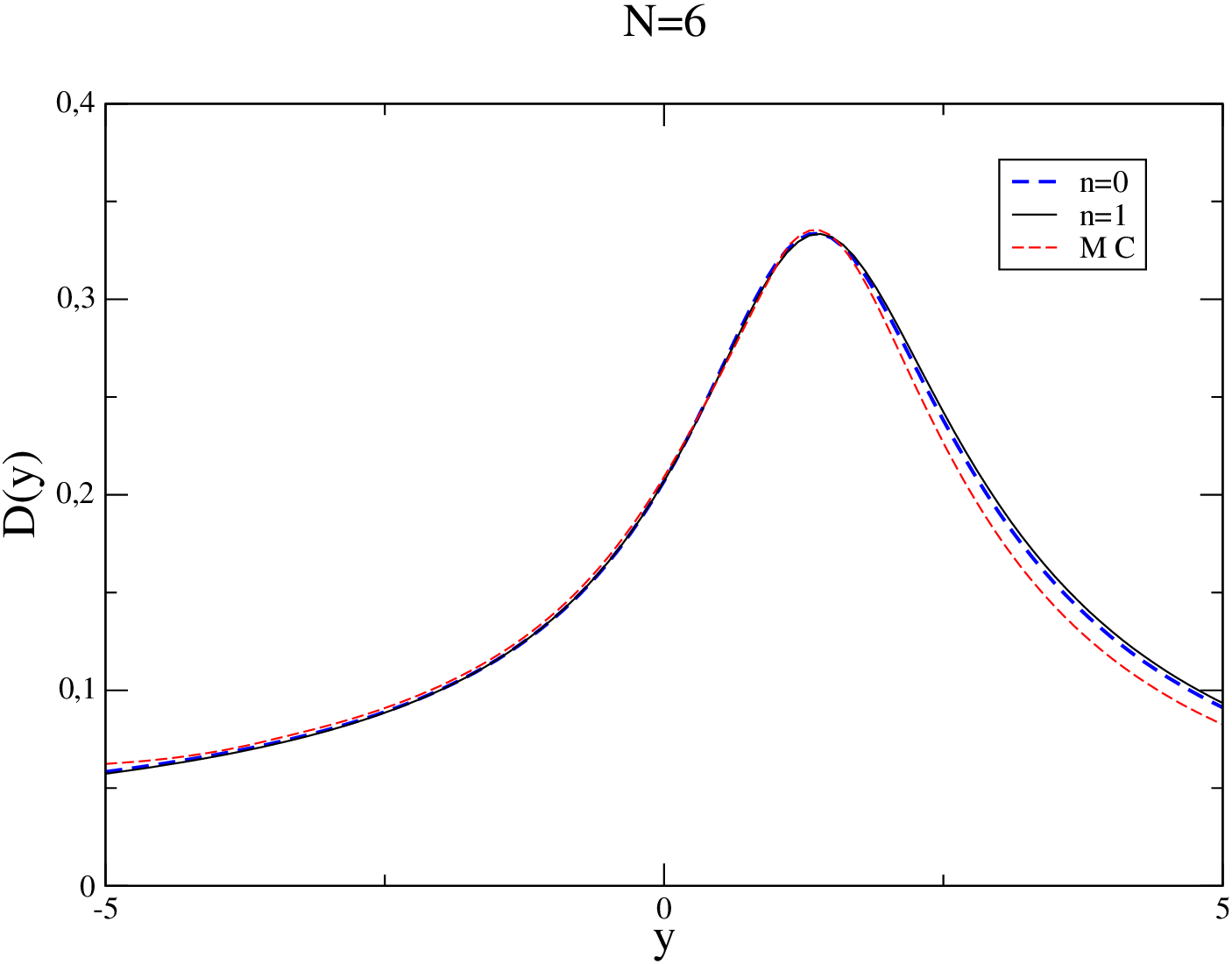}
\end{minipage}
\\
\caption{The scaling function $D(y)$ for the models $N=5$ and $N=6$. For the $O(6)$ model we show, for comparison, the results of Ref. \protect{\cite{MCO6}}.}
\label{figDy}
\end{figure}

As explained in section \ref{approx_eq}, in order to compute the scaling equation of state and universal amplitude ratios, we need the input values of the critical exponents and universal couplings. We chose those final quantities by comparing the values obtained by resumming the fixed-dimension and the $\epsilon$-expansion series. For the $O(5)$ model our input values are $\eta=0.031(3)$, $\nu=0.764(4)$, $g_4=15.74(2)$, $r_6=1.71(2)$, $r_8=-0.3(5)$. For the $O(6)$ model they are $\eta=0.029(3)$, $\nu=0.789(5)$, $g_4=14.43(4)$, $r_6=1.65(3)$, $r_8=-0.7(4)$. Since we have $r_6$ and $r_8$, we can implement the approximation scheme up to the step $n=1$. The results for universal amplitude ratios and other universal quantities are presented in table \ref{results56}, along with a comparison with the $1/N$ expansion. The errors reported are due to the variation of the input parameters in their error intervals. Inspecting table \ref{results56}, we see that all the results obtained in the $n=1$ scheme are consistent, within the errors, with the results for the $n=0$ case. In particular, for the $N=5$ model, the central values for the $n=1$ scheme are very close to those for $n=0$: this is due to the fact that the predicted $r_8$ in the $n=0$ case matches the input value used in the $n=1$ scheme. Moreover, the condition (\ref{consistency}) is satisfied. The equation of state for the $O(6)$ model has been obtained by means of Monte Carlo simulation in Ref. \cite{MCO6}. The comparison should be done with some cautions, since the critical exponents in Ref. \cite{MCO6} are different from the ones we used in this work (see table \ref{exp56}). We report the results: $R_\chi = 1.09(1)$, $y_{max} = 1.34(5)$. Moreover, from the fits of Ref. \cite{MCO6} other universal quantities can be obtained: $R_4^+ = 7.6(2)$, $F_0^\infty = 0.0233(9)$, $c_f = 2.1(2)$, $P_m = 1.13(2)$, $P_c = 0.36(6)$. Comparing with our results shown in table \ref{results56}, we see that there is a slight difference among the universal amplitude ratios involving the high-temperature phase. The scaling functions $f(x)$, $F(z)$, $E(y)$ and $D(y)$ are displayed, respectively, in figures \ref{figfx}, \ref{figFz}, \ref{figEy} and \ref{figDy}. In figures \ref{figEy} and \ref{figDy} we also show a comparison with the Monte Carlo fits of Ref. \cite{MCO6}. As can be seen, the curves $n=0$ and $n=1$ are very similar. We observe a little difference between our curves and the Monte Carlo fits, in particular for large negative $y$ in $E(y)$ and for large positive $y$ in $D(y)$. This, again, can be interpreted as coming from the different critical exponents used. In fact, for $y\rightarrow -\infty$, $E(y)\simeq (-y)^\beta$: the discrepancies at large negative $y$ are caused by the $\beta=0.425(2)$ exponent used in Ref. \cite{MCO6}, obtained by their MC simulations, which is larger than our $\beta=0.406(3)$, obtained by field-theory methods. For $y\rightarrow +\infty$, $D(y)\simeq R_\chi y^{-\gamma}$. While our universal amplitude ratio $R_\chi$ is compatible with that of Ref. \cite{MCO6}, the critical exponent $\gamma=1.555(10)$ used in this work is smaller than the one used in Ref. \cite{MCO6}, $\gamma=1.604(6)$.
Finally, we mention that the coefficient $F_0^\infty$ appears to be very
precise. This is not trivial, since the series (\ref{expF}) has a
finite radius of convergence: starting from the small-$z$ (high-temperature) expansion, the approximation scheme provides an analytic
continuation to the low-temperature phase.
On the other hand, we see that the $1/N$ expansion provides values
that differ significantly from the results obtained.

\begin{table*}
\caption{Universal amplitude ratios and other quantities for the models $N=5$ and $N=6$.}
\footnotesize
\begin{tabular}{|lccc||ccc|}
\hline
\label{results56}
& & $N=5$ & & & $N=6$ &\\
\hline
& n=0 & n=1 & 1/N & n=0 & n=1 & 1/N \\
\hline
$\rho$ & 2.16(1) & 2.16(9) & - & 2.15(2) & 2.12(7) & - \\
$\theta_0^2$ & 2.75(6) & 2.7(2) & - & 2.56(7) & 2.5(1) & - \\
$c_1$ & - & 0.0006(234) & - & - & 0.008(19) & - \\
\hline
$r_8$ & -0.3(1) & -0.3(5)* & -13.46 & -0.5(1) & -0.7(4)* & -11.22 \\
$r_{10}$ & 2.7(8) & 3(8) & 281.4 & 5(1) & 8(8) & 234.5 \\
$U_0$ & 2.21(3) & 2.2(2) & 1.47 & 2.39(5) & 2.5(2) & 1.47 \\
$R_\alpha$ & 4.1(1) & 4.2(6) & 0.77 & 3.8(2) & 4.0(5) & 0.72 \\
$R_c^+$ & 0.279(9) & 0.28(2) & 0.595 & 0.33(1) & 0.33(2) & 0.595 \\
$R_4^+$ & 8.25(9) & 8.3(5) & 1.18 & 8.3(1) & 8.4(4) & 2.99 \\
$R_\chi$ & 1.20(2) & 1.2(1) & 0.62 & 1.12(3) & 1.15(9) & 0.681 \\
$R_\xi^+$ & 0.527(5) & 0.527(6) & 0.6185 & 0.575(6) & 0.575(6) & 0.6573 \\
$F_0^\infty$ & 0.0214(5) & 0.0214(7) & 0.01961 & 0.0196(5) & 0.0195(6) & 0.01750 \\
$f_0^1$ & 1.48(2) & 1.48(9) & 1.782 & 1.57(2) & 1.55(6) & 1.819 \\
$f_0^2$ & 0.33(2) & 0.33(6) & 0.620 & 0.41(2) & 0.40(5) & 0.684 \\
$f_0^3$ & -0.064(2) & -0.06(2) & -0.129 & -0.073(2) & -0.07(1) & -0.107 \\
$c_f$ & 3.9(4) & 4(3) & 1.3 & 2.7(2) & 3(1) & 1.3 \\
$P_m$ & 1.153(6) & 1.15(3) & 1.029 & 1.133(6) & 1.14(2) & 1.040 \\
$P_c$ & 0.330(2) & 0.330(4) & 0.4088 & 0.319(3) & 0.318(5) & 0.3741 \\
$R_p$ & 2.069(4) & 2.069(9) & 1.6593 & 2.088(5) & 2.091(9) & 1.7827 \\
$z_{max}$ & 1.311(3) & 1.311(6) & 1.0194 & 1.326(4) & 1.327(7) & 1.1077 \\
$x_{max}$ & 7.3(2) & 7.3(6) & 5.22 & 6.7(2) & 6.9(4) & 5.18 \\
$y_{max}$ & 1.44(2) & 1.4(1) & 1.089 & 1.36(2) & 1.38(7) & 1.107 \\
\hline
\end{tabular}\\
$*$ Input values
\end{table*}

\section{Critical exponents and universal couplings for large $N$}
\label{criticalNlarge}
We present in table \ref{expNlarge} the results for critical exponents and
universal couplings for the models $N=32$ and $N=64$, obtained by an analysis of the fixed-dimension perturbative series \cite{Baker_series,sokolov_exp,sokolov_r2n} and of the $\epsilon$-expansion series \cite{Gorishnii,Kleinert,Pelissetto:1997gk,Pelissetto:1998sk,potential2000}.
In fixed-dimension we resum the series for the critical exponents $\eta(g)$,
$\eta_t(g)$ and for the beta function $\beta(g)$ up to six loops, and the
series for $r_6(g)$ and $r_8(g)$ up to four and three loops respectively.
In $\epsilon$-expansion we resum the series for $\eta(\epsilon)$ and
$\eta_2(\epsilon)$ up to five loops and the series for $r_6(\epsilon)$,
$r_8(\epsilon)$ and $g_4(\epsilon)$ up to three loops.
We also show a comparison with other determinations available in literature and with the large-$N$ expansion. The resummation of universal couplings in \mbox{$\epsilon$-expansion} has been constrained with the exact results available in dimensions $d=0$ and $d=1$. The resummation is performed with the conformal mapping method; see appendix \ref{resummation} for details.

\begin{table*}
\caption{Critical exponents, universal couplings and the universal quantity $g_4$ for the models $N=32$ and $N=64$ obtained by resummation of the fixed-dimension series \protect{\cite{Baker_series,sokolov_exp,sokolov_r2n}} (first row) and $\epsilon$-expansion series \protect{\cite{Gorishnii,Kleinert,Pelissetto:1997gk,Pelissetto:1998sk,potential2000}} (second row). A comparison with other determinations is shown.}
\footnotesize
\begin{tabular}{|cl|c|c|c|c|c|c|}
\hline
\label{expNlarge}
$N$ &                & $g_4$        & $\eta$ & $\nu$ & $r_6$ & $r_8$ \\
\hline
$32$& fixed-dim      & $4.2295(5)$  & $0.00901(1)$ & $0.9584(2)$ & $1.1106(7)$ & $-1.3(2)$ \\
    & $\epsilon$-exp & $4.209(3)$   & $0.0088(2)$  & $0.972(3)$  & $1.112(6)$  & $-1.244(7)$ \\
    &                & $^a 4.230$   & $^a 0.009$   & $^a 0.958$  & $^b 1.111$ & $^b -1.23$ \\
    &                & $^c 4.22(1)$ &              &             & $^c 1.110(9)$ & $^c -1.24(2)$\\
    &                &              & $^d 0.00844$ & $^d 0.966$  & $^d 1.152$ & $^d -2.1$\\
\hline
$64$& fixed-dim      & $2.23177(4)$ & $0.00467(2)$ & $0.9803(2)$ & $0.9819(4)$ & $-0.86(9)$ \\
    & $\epsilon$-exp & $2.2280(3)$  & $0.00404(5)$ & $0.982(6)$ & $0.9846(3)$ & $-0.785(2)$ \\
    &                &              & $^d 0.00422$ & $^d 0.9831$ & $^d 0.9929$ & $^d -1.052$\\
\hline
\end{tabular}\\
$^a$\cite{sokolov_exp} FT $d=3$ exp,
$^b$\cite{sokolov_r2n} FT $d=3$ exp\\
$^c$\cite{potential2000} FT $\epsilon$-exp constrained\\
$^d$Large-$N$ expansion to order $O(1/N)$, see appendix \ref{largeN}\\
\end{table*}

When $N$ is large enough, the coefficients of the series become
small. Thus, we expect that the errors provided by the resummation
technique become smaller. However, as we state in appendix \ref{resummation}, the errors
obtained by the analysis of the series don't have a statistical
meaning: they are the variation of the quantity at hand, when the
resummation parameters are varied. We believe that the errors provided
by our procedure are substantially underestimated when $N$
is large. In order to get the input values for the approximate equation of state with reasonable errors, we must again compare the
results obtained with the fixed-dimension with those obtained with the $\epsilon$-expansion. For the $O(32)$ model our input values are \mbox{$\eta=0.0089(3)$}, \mbox{$\nu=0.965(8)$}, $g_4=4.22(1)$, \mbox{$r_6=1.112(6)$}, $r_8=-1.24(2)$. For the $O(64)$ model they are $\eta=0.0044(4)$, $\nu=0.981(4)$, $g_4=2.230(2)$, $r_6=0.983(2)$, $r_8=-0.79(1)$.

In table \ref{results3264} we present the results for various universal quantities, for the models $N=32$ and $N=64$, along with a comparison with the $1/N$ expansion. The errors reported are due to the variation of the input parameters in their error intervals. Inspecting table \ref{results3264}, we see that, for both models, all the values obtained in the $n=0$ scheme are consistent, within the errors, with those obtained in the scheme $n=1$. In particular, in many cases the central values, in the two approximation steps, are very close. Regarding the $1/N$ expansion, we note that, except for the coefficients $r_{2n}$, even in the $N=32$ model there is a good agreement with our universal quantities and those coming from $1/N$ expansion of the equation of state. The agreement is even better for the $N=64$ model: in this case almost all the quantities calculated in the $n=1$ approximation scheme are compatible, within the errors, with those obtained in the $1/N$ expansion.

\begin{table*}
\caption{Universal amplitude ratios and other quantities for the models $N=32$ and $N=64$.}
\footnotesize
\begin{tabular}{|lccc||ccc|}
\hline
\label{results3264}
& & $N=32$ & & & $N=64$ &\\
\hline
& n=0 & n=1 & 1/N & n=0 & n=1 & 1/N \\
\hline
$\rho$ & 1.9(1) & 1.90(7) & - & 1.91(10) & 1.76(5) & - \\
$\theta_0^2$ & 1.51(9) & 1.51(1) & - & 1.48(8) & 1.406(7) & - \\
$c_1$ & - & -0.0001(209) & - & - & 0.02(2) & - \\
\hline
$r_8$ & -1.2(3) & -1.24(2)* & -2.104 & -0.7(1) & -0.79(1)* & -1.052 \\
$r_{10}$ & 19(7) & 19(3) & 44 & 10(3) & 15(2) & 22 \\
$U_0$ & 1.5(3) & 1.5(5) & 1.47 & 1.4(2) & 1.5(4) & 1.47 \\
$R_\alpha$ & 0.5(3) & 0.5(6) & 0.51 & 0.4(2) & 0.6(4) & 0.49 \\
$R_c^+$ & 0.55(3) & 0.55(7) & 0.595 & 0.55(3) & 0.58(5) & 0.595 \\
$R_4^+$ & 10.4(1) & 10.42(4) & 10.310 & 11.10(7) & 11.14(3) & 11.155 \\
$R_\chi$ & 0.939(2) & 0.94(1) & 0.940 & 0.965(1) & 0.968(4) & 0.9701 \\
$R_\xi^+$ & 1.11(1) & 1.11(4) & 1.148 & 1.40(2) & 1.42(4) & 1.447 \\
$F_0^\infty$ & 0.0092(2) & 0.0092(2) & 0.00892 & 0.0081(1) & 0.00805(8) & 0.007934 \\
$f_0^1$ & 1.97(1) & 1.97(2) & 1.966 & 1.984(5) & 1.983(7) & 1.9830 \\
$f_0^2$ & 0.94(2) & 0.94(3) & 0.941 & 0.971(9) & 0.97(1) & 0.970 \\
$f_0^3$ & -0.020(6) & -0.020(7) & -0.0201 & -0.011(3) & -0.011(3) & -0.0100 \\
$c_f$ & 1.04(2) & 1.04(3) & 1.051 & 1.020(7) & 1.02(1) & 1.026 \\
$P_m$ & 1.0800(5) & 1.080(3) & 1.0822 & 1.0853(2) & 1.0860(9) & 1.08710 \\
$P_c$ & 0.233(3) & 0.233(1) & 0.2334 & 0.218(1) & 0.2177(6) & 0.21717 \\
$R_p$ & 2.295(9) & 2.295(4) & 2.2843 & 2.340(5) & 2.343(2) & 2.3421 \\
$z_{max}$ & 1.475(5) & 1.475(2) & 1.4664 & 1.507(3) & 1.509(1) & 1.5078 \\
$x_{max}$ & 5.00(5) & 5.00(7) & 5.034 & 5.01(2) & 5.01(3) & 5.017 \\
$y_{max}$ & 1.171(2) & 1.171(8) & 1.1764 & 1.181(1) & 1.182(3) & 1.1844 \\
\hline
\end{tabular}\\
$*$ Input values
\end{table*}

\begin{ack}
We are grateful to Ettore Vicari for his support, useful discussions and for reading carefully the manuscript. We thank Pasquale Calabrese for useful discussions. We thank Sven Holtmann and Thomas Schulze for sending us the plots.
\end{ack}

\appendix

\section{Resummation of perturbative series}
\label{resummation}

We have calculated the critical exponents and the coefficients $r_6$ and $r_8$ for the $O(N)$ model in three dimensions using the perturbative series from field theory available in the literature; these quantities are the input parameters for the determination of the critical equation of state and of the universal amplitude ratios, as explained in section \ref{approx_eq}.

Since perturbative expansions are divergent (asymptotic), a resummation procedure is needed to obtain reasonable numerical estimates. Consider a generic quantity $S(g)$ that has a perturbative expansion
\begin{equation}
S(g)= \sum s_k g^k
\end{equation}
in the renormalized coupling constant $g$, corresponding to the coupling term of the $\phi^4$-Hamiltonian we are considering. We use a normalization for $g$ such that the beta function may be written as $\beta(g)= -g+g^2+O(g^3)$ \cite{sokolov_exp}. The large order behavior of the coefficients is generally given by
\begin{equation}
s_k \sim k! (-a)^k k^c \left[ 1+O(k^{-1}) \right].
\end{equation}
Borel summability for $g>0$ (proved in fixed dimensions $d<4$ \cite{eckmann} for $O(N)$ symmetric models) requires $a>0$. The constant $a$ does not depend on the specific observable, unlike the constant $c$. From instanton calculations (see rfs. \cite{ZJbook,Lipatov:1976ny,Brezin:1976vw,parisi}) one can determine
\begin{equation}
a=1.32996798/(8+N),
\label{costa}
\end{equation}
for the $O(N)$ model in three dimensions. We then introduce the Borel-Leroy transform $B(t)$ of $S(g)$
\begin{equation}
S(g)=\int_0^\infty \mathrm{d}t \, t^b e^{-t}B(gt),
\label{integrale}
\end{equation}
where $b$ is a resummation parameter. Its expansion is
\begin{equation}
B_{\mathrm{exp}}(t)=\sum_k \frac{s_k}{\Gamma(k+b+1)} t^k.
\label{expansion}
\end{equation}
This expansion is now convergent in the disk of the complex plane: \mbox{$|t|<1/a$}, where $a$ is the constant (\ref{costa}) that characterizes the large order behavior of $S(g)$. In fact the singularity of $B(t)$ that is nearest to the origin is in \mbox{$t_s=-1/a$}. However, in order to reconstruct the function $S(g)$ from the integral in (\ref{integrale}), we need to perform an analytic continuation of the expansion (\ref{expansion}). For example one can use Pad\'e approximant to the series (\ref{expansion}): this is the Pad\'e-Borel method. We have used instead a more refined procedure, the conformal mapping method, based on the knowledge of the constant $a$: following Ref. \cite{LeGuillou:ju}, we perform an Euler transformation 
\begin{equation}
y(t)=\frac{\sqrt{1+at}-1}{\sqrt{1+at}+1},
\end{equation} 
and rewrite the Borel transform as
\begin{equation}
B(t)=\frac{1}{\left[1-y(t)\right]^\alpha} \sum_k b_k y(t)^k,
\label{bdit}
\end{equation}
where $\alpha$ is another resummation parameter and the coefficients $b_k$ are determined by the requirement that the expansion of $B(t)$ in eq. (\ref{bdit}) in powers of $t$ is equal to the series (\ref{expansion}). If all singularities of $B(t)$ belong to the real interval $\left] -\infty , -t_c\right]$, the expansion converges everywhere in the complex $t$-plane except on the negative axis for $t<-t_c$.
A good numerical estimate of $S(g)$ is therefore given by
\begin{equation}
S(g)\simeq \sum_{k=0}^p b_k \int_0^\infty \mathrm{d}t \, t^b e^{-t} \frac{y(gt)^k}{\left[1-y(gt)\right]^\alpha},
\end{equation}
where now the sum goes until $k=p$, since obviously we know only a finite number of terms of the perturbative series.

We have also analyzed the $\epsilon$-expansion of the critical exponents and of the coefficients $r_6$ and $r_8$, using the same conformal mapping method for the resummation of the series in $\epsilon$ (where $\epsilon=4-d$) and setting then $\epsilon =1$. The constant $a$, which characterizes the singularity of the Borel transform, is given by
\begin{equation}
a=\frac{3}{8+N},
\end{equation} 
for the $\epsilon$-expansions in the $O(N)$ model \cite{Lipatov:1976ny,Brezin:1976vw}.

Before resumming any series $S(g)$ (or $S(\epsilon)$), we divided it by an opportune power of $g$ (or $\epsilon$), so that the new series starts with a nonzero constant.

In the framework of fixed dimension expansion ($d=3$), we numerically resummed the series for:
\begin{itemize}
\item $\beta(g)$ up to six loops \cite{Baker_series,sokolov_exp}; we determined the IR fixed point $g^*$: $\beta(g^*)=0$.
We recall that the universal quantity $g_4$, defined in table \ref{amplitudes_def} and reported in tables \ref{exp56}-\ref{expNlarge}, is related to $g^*$ by: $g_4=48 \pi g^* /(N+8)$, for the three-dimensional $O(N)$ model.
\item $\eta(g)$ and $\eta_t(g)$ up to six loops \cite{Baker_series,sokolov_exp}, using the value of $g^*$. The other critical exponents were determined through the hyperscaling relations among them.
\item $r_6(g)$ and $r_8(g)$ up to four and three loops respectively \cite{sokolov_r2n}. 
\end{itemize}
As to the $\epsilon$-expansion we resummed the series for:
\begin{itemize}
\item $\eta(\epsilon)$ and $\eta_2(\epsilon)$ up to five loops \cite{Gorishnii,Kleinert}. The other critical exponents were determined through the hyperscaling relations among them.
\item $g_4(\epsilon)$, $r_6(\epsilon)$ and $r_8(\epsilon)$ up to three loops \cite{Pelissetto:1997gk,Pelissetto:1998sk,potential2000}.
\end{itemize}
The values of $g_4(\epsilon)$ and $r_{2j}(\epsilon)$ are known exactly in dimensions $d=0$ and $d=1$; therefore we have also performed for these quantities constrained analyses using the same method described in \cite{Pelissetto:1997gk,Pelissetto:1998sk,potential2000} and references therein, which should improve the final results under the assumption of a sufficiently large analytic domain in $\epsilon$ for the quantity to be resummed.

Suppose that the exact values $R_{\mathrm{ex}}(\epsilon_1), \ldots R_{\mathrm{ex}}(\epsilon_k)$ of the generic quantity $R(\epsilon)$ are known for a set of dimensions $\epsilon_1 , \ldots \epsilon_k$. Then define
\begin{equation}
Q(\epsilon)= \sum_{i=1}^{k} \left[ \frac{R_{\mathrm{ex}}(\epsilon_i)}{\epsilon - \epsilon_i}
 \prod_{j=1,j \neq i}^k \left( \epsilon_i - \epsilon_j \right)^{-1}\right]
\end{equation}
and
\begin{equation}
S(\epsilon)=\frac{R(\epsilon)}{\prod_{i=1}^k \left( \epsilon - \epsilon_i \right)} - Q(\epsilon)
\end{equation}
and finally
\begin{equation}
R_{\mathrm{imp}}(\epsilon)= \left[ Q(\epsilon) + S(\epsilon) \right]\prod_{i=1}^k \left( \epsilon - \epsilon_i \right)
\end{equation} 
The resummation procedure is applied to $S(\epsilon)$ and the final estimate is obtained by computing $R_{\mathrm{imp}}(\epsilon=1)$. When the polynomial interpolation through the values $R_{\mathrm{ex}}(\epsilon_1), \ldots R_{\mathrm{ex}}(\epsilon_k)$ is a good approximation for $R(\epsilon)$, one expects that the $\epsilon$ series which gives the deviation has smaller coefficients than the original one. In fact one can check that, in the cases considered, the coefficients of $S(\epsilon)$ decrease with $k$. Consequently also the errors in the resummation should be smaller.
   
Any resummed quantity depends on the arbitrary parameters $\alpha$ and $b$: $S_p(\alpha,b)$, where $p$ is the number of known terms in the original series. The parameters $\alpha$ and $b$ may be used to optimize the resummation, see e.g. refs \cite{LeGuillou:ju,Guida:1998bx,Carmona:1999rm}. We shall now describe the algorithm we used; the idea is to choose the values of $\alpha$ and $b$ providing the fastest convergence of the results in the number of loops. 

We tried to minimize numerically the quantity
\begin{equation}
\left(\frac{S_p(\alpha,b)}{S_{p-1}(\alpha,b)}-1 \right)^2,
\label{min}
\end{equation}
in a suitable range of parameters: $-3/2 \lesssim \alpha \lesssim 2$ (for $N=5$, $6$), $-2 \lesssim \alpha \lesssim 5$ (for $N=32$, $64$) and $0 \lesssim b \lesssim 20$ (for all the $N$ considered). $S_{p-1}$ is the quantity resummed using only $p-1$ terms of the original series. When we found only one minimum of (\ref{min}), we considered the positions of the minimum $(\alpha_{\mathrm{opt}}, b_{\mathrm{opt}})$ as the best resummation parameters.  
However in many cases, we found a whole line of zeroes for $S_p - S_{p-1}$ in the plane $(\alpha,b)$, that is a line of minima for (\ref{min}). In those cases we chose the optimal parameters $(\alpha_{\mathrm{opt}}, b_{\mathrm{opt}})$ by minimizing the quantity
\begin{equation}
\left(\frac{S_{p-1}(\alpha,b)}{S_{p-2}(\alpha,b)}-1 \right)^2,
\label{min2}  
\end{equation}
along the line of zeroes for $S_p - S_{p-1}$.

Once we have determined the best resummation parameters, we considered different sets of approximants for the quantity to be resummed \cite{Carmona:1999rm}, obtained by varying the parameters in an opportune range $\alpha \in \left[ \alpha_{\mathrm{opt}}- \Delta \alpha \right.$, $\left. \alpha_{\mathrm{opt}} + \Delta \alpha \right]$ and $b \in \left[ b_{\mathrm{opt}}- \Delta b, b_{\mathrm{opt}} + \Delta b \right]$. 
One can take as the final estimate the average of $S_p(\alpha,b)$ with $\alpha$ and $b$ in this range (we considered half-integers $\alpha$ and integers $b$, after approximating $\alpha_{\mathrm{opt}}$ and $b_{\mathrm{opt}}$ to the nearest half-integer or integer, respectively), while the error is indicated by the variance of $S_p(\alpha,b)$.
It is not so clear how to determine the width of the intervals $\Delta \alpha$ and $\Delta b$ \cite{Carmona:1999rm}: in general the average of $S_p$ is stable, within the quoted errors, while the variance strongly depends on the choice one is making. In order to obtain reasonable errors, we compared our results with those obtained by other authors in the case $N=4$, and determined $\Delta \alpha$ and $\Delta b$ so as to reproduce their errors. We have chosen $\Delta \alpha=2$ and $\Delta b=3$; moreover we quote the error in the final results as two standard deviations.
In this way we obtain for $g^*$ with $N=4$ the value $g^*=1.375(4)$, which agrees with the result quoted by R.~Guida and J.~Zinn-Justin in Ref. \cite{Guida:1998bx}: $g^*=1.377(5)$.   

For series in fixed dimension depending on $g^*$ we summed the squared errors due to the uncertainties of $\alpha$, $b$, and of $g^*$; the latter was estimated by $(S_p(\alpha_{\mathrm{opt}},b_{\mathrm{opt}},g^* + \Delta g^*)-S_p(\alpha_{\mathrm{opt}},b_{\mathrm{opt}},g^* - \Delta g^*))/2$.

This algorithm is ad hoc and in some way arbitrary, but provides estimates consistent with each other. Moreover our results are in substantial agreement with those of other authors in the case $N=4$ \cite{Guida:1998bx}. Therefore we consider our final values and error bars reliable, although we should be cautious in giving them the standard statistical meaning.

\section{Large $N$ behavior of the equation of state}
\label{largeN}
We consider now the large-$N$ behavior of the critical exponents and of the equation of state for the three dimensional $O(N)$ model; our aim is to compare the results from the $1/N$ expansion with the numerical estimates for critical quantities obtained through resummation procedures and through the parametric approximation of the equation of state.

The longest $1/N$ series for critical exponents available in literature are \cite{OkabeN2,etaN3}:
\begin{align}
\label{exponenta}
\displaystyle
&\alpha=-1 + {\frac{32}{N\,{{\pi }^2}}} - \frac{96}{N^2\pi^4}\left(\frac{112}{27}-\pi^2\right), \\[1em]
\label{exponentbeta}
\displaystyle
&\beta={\frac{1}{2}} - {\frac{4}{N\,{{\pi }^2}}} + \frac{16}{N^2\pi^4}\left(\frac{8}{3}-\pi^2\right), \\[1em]
\label{exponentgamma}
\displaystyle
&\gamma=2 - {\frac{24}{N\,{{\pi }^2}}} + \frac{64}{N^2\pi^4}\left(\frac{44}{9}-\pi^2\right), \\[1em]
\label{exponentdelta}
\displaystyle
&\delta=5 - {\frac{16}{N\,{{\pi }^2}}} + \frac{1408}{N^2 9\pi^4} - \frac{1024}{N^3 27\pi^6}\left[\frac{27}{2}\pi^2\log{2}+\frac{81}{8}\psi^{''}\left(\frac{1}{2}\right)-\frac{61}{8}\pi^2-\frac{683}{6}\right], \\[1em]
\label{exponenteta}
\displaystyle
&\eta={\frac{8}{N\,3{{\pi }^2}}}-\frac{512}{N^2 27\pi^4} + \frac{512}{N^3 27\pi^6}\left[\frac{9}{2}\pi^2\log{2}+\frac{27}{8}\psi^{''}\left(\frac{1}{2}\right)-\frac{61}{24}\pi^2-\frac{797}{18}\right], \\[1em]
\label{exponentnu}
\displaystyle
&\nu=1 - {\frac{32}{N\,3{{\pi }^2}}} + \frac{32}{N^2\pi^4}\left(\frac{112}{27}-\pi^2\right),
\end{align}
where $\psi(x)$ is the logarithmic derivative of the $\Gamma$ function. All the equalities (\ref{exponenta}-\ref{exponentnu}) are understood to be true up to higher orders in $1/N$. Since, in order to compute the large-$N$ expansion of the various universal quantities, we used the equation of state up to the order $1/N$, we have, for consistence, limited ourself to the order $1/N$ in the critical exponents.
The large-$N$ limit of the equation of state and the $1/N$ corrections were calculated in Ref. \cite{largeNeq}; since we will use them, we report their results in the case of three spatial dimensions
\begin{equation}
f(x)={{\left( x+1 \right) }^2} + \frac{2}{N} \left( x+1 \right) 
   \left[ g(x) - \left( x+1 \right) \,g(0)  - {\frac{12\,\left( x+1 \right) \,\log (x+1)}{{{\pi }^2}}} \right],
\label{effe}
\end{equation}
where $g(x)$ is given by
\begin{equation}
g(x)=\frac{1}{\pi^2} \int _{0}^{\infty}{\frac{1}{{\frac{1}{2\,\pi \,\left( 1 + x \right) }} + {k^2}\,i(k,0)}} + 
     {\frac{-1 - 2\,\pi \,\left( 1 + x \right) \,i(k,1) + {\frac{2\,{k^2}\,j(k,1)}{i(k,1)}}}
       {{\frac{1}{2\,\pi \,\left( 1 + x \right) }} + \left( 1 + {k^2} \right) \,i(k,1)}}\,dk,
\label{gi}
\end{equation}
and the functions $i(k,r)$ and $j(k,r)$ are defined as in Ref. \cite{largeNeq}
\begin{equation}
\begin{array}{l}
i(k,r)=\displaystyle
\frac{1}{4 \pi k} \, \arctan \! \left( \frac{k}{2\, \sqrt{r}} \right),\\[2em]
j(k,r)=\displaystyle
{\frac{-3\,{\sqrt{r}}}
   {8\,\pi \,\left( {k^2} + r \right) \,
     \left( {k^2} + 4\,r \right) }}.
\label{ij}
\end{array}
\end{equation}
The function $f(x)$ in (\ref{effe}) has the right normalization: $f(-1)=0$ and \mbox{$f(0)=1$}. 
Starting from these equations, we derived the large-$N$ expansion of universal amplitude ratios and of other quantities.

The functions $f(x)$ and $g(x)$ are regular in $x=0$; therefore it is easy to compute the coefficients $f^0_i$
\begin{align}
f^0_1 &= \displaystyle 2 + \frac{1}{N} \left(-{\frac{24}{{{\pi }^2}}} - 2\,g(0) + 2\,g'(0)\right),\\[1em]
f^0_2 &= \displaystyle 1 + \frac{1}{N} \left(- {\frac{36}{{{\pi }^2}}} - 2\,g(0) + 2\,g'(0) + g''(0)\right),\\[1em]
f^0_3 &= \displaystyle \frac{1}{N} \left(-{\frac{8}{{{\pi }^2}}} + g''(0) + {\frac{g^{(3)}(0)}{3}}\right).
\end{align}
The values of $g(x)$ and its derivatives in $x=0$ can be obtained from the definition (\ref{gi}) by calculating numerically the integrals in $k$.
In the opposite limit $x\rightarrow \infty$, we wrote the behavior of $g(x)$ as
\begin{equation}
g(x)=-(x+1)+\frac{8}{\pi^2}\log(x+1)+\frac{8}{\pi^2}\log \frac{\pi}{4} +\int_0^\infty dk \, R(k,x),
\label{inf}
\end{equation}
where the function $R(k,x)$ is
\begin{equation}
R(k,x)=
{\frac{8}{\left( 1 + k \right) \,{{\pi }^2}}} + 
  {\frac{2\,\left( 1 + x \right) }
    {\left( 1 + {k^2} \right) \,\pi }} + 
  {\frac{-1 - 2\,\pi \,\left( 1 + x \right) \,i(k,1) + 
      {\frac{2\,{k^2}\,j(k,1)}{i(k,1)}}}{{{\pi }^2}\,
      \left( {\frac{1}{2\,\pi \,\left( 1 + x \right) }} + 
        \left( 1 + {k^2} \right) \,i(k,1) \right) }}.
\end{equation}
The last term in (\ref{inf}) can be expanded in powers of $1/(x+1)$
\begin{equation}
\int_0^\infty dk \, R(k,x)= g_0 + \frac{g_1}{x+1} + \frac{g_2}{(x+1)^2} + \frac{g_3}{(x+1)^3} 
+ O \left( \frac{1}{(x+1)^4} \right).
\end{equation}
The coefficients $g_i$ can be obtained by Taylor expanding the integrand and calculating numerically the integrals in $k$. Then one can find the coefficients $f^\infty_i$
\begin{align}
f^\infty_0 &= \displaystyle
1 - \frac{2}{N}\,\left( 1 + g(0) \right),\\[1em]
f^\infty_1 &= \displaystyle
2 + \frac{2}{N} \left( -2 +  g_0 - {\frac{12}{{{\pi }^2}}} -
  2\,g(0) + {\frac{8\,\log ({\frac{\pi }{4}})}
    {{{\pi }^2}}} \right),\\[1em]
f^\infty_2 &= \displaystyle
1 + \frac{2}{N}\,\left( -1 + g_0 + g_1 -
     {\frac{10}{{{\pi }^2}}} - g(0) +
     {\frac{8\,\log ({\frac{\pi }{4}})}{{{\pi }^2}}}
     \right),\\[1em]
f^\infty_3 &= \displaystyle
\frac{2 g_2}{N}.
\end{align}
Using these results and the definitions of universal amplitude ratios, one can find:
\begin{align}
R_4^+ &= \displaystyle
12+ \frac{12}{N}\left({\frac{-12 + g_0\,{{\pi }^2} + 
     8\,\log ({\frac{\pi }{4}})}{{{\pi }^2}}}\right),\\[1em]
R_{\chi} &= \displaystyle     
1 + \frac{2}{N}\,\left( 1 + g(0) \right),\\[1em]
F^\infty_0 &= \displaystyle
\frac{1}{144}+ \frac{1}{N} \, {\frac{{{\pi }^2}\,\left( 1 - g_0 + 
        g(0) \right)  + 
     4\,\left( 3 + \log (192) - 2\,\log (\pi ) \right) }
     {72\,{{\pi }^2}}},\\[1em]
r_6 &= \displaystyle
\frac{5}{6} + \frac{1}{N} \, {\frac{5\,
      \left( 2 + g_1\,{{\pi }^2} \right) }{3\,
      {{\pi }^2}}},\\[1em]
r_8 &= \displaystyle
\frac{1}{N} \,\frac{35\,g_2}{6},\\[1em]
r_{10} &= \displaystyle     
\frac{1}{N} \,\frac{-35\,\left( 2 + 6\,g_2\,{{\pi }^2} - 
       6\,g_3\,{{\pi }^2} \right) }{6\,
     {{\pi }^2}}.
\end{align}
All these equalities are understood to be true up to order $1/N$.

To derive the expressions of universal ratios involving the specific heat amplitudes, we need the effective potential (Helmholtz free energy), proportional to the universal function $A(x)$. Using the notations of Ref. \cite{review}, we have 
\begin{equation}
f(x)=A(x)-\frac{x}{d\nu}A'(x).
\label{diff}
\end{equation}
$A(x)$ is the solution of this differential equation which is analytic in $x=0$. 
Solving (\ref{diff}) using the expressions (\ref{exponenta}-\ref{gi}) for $f(x)$ and critical exponents up to $1/N$ terms, and then taking the limit $N\rightarrow \infty$, we obtained for $A(x)$
\begin{equation}
A(x)=1+3x+3x^2-\frac{\pi^2}{32}\left( -\frac{24}{\pi^2} +3g''(0)+g'''(0) \right)x^3.
\label{a}
\end{equation}
Note that the knowledge of $f(x)$ and $\nu$ up to the order $1/N$ allows to determine only the leading term of the $1/N$ expansion of $A(x)$. 
This is related to the fact that, in dimension $d=3$, the critical exponent $\alpha$ tends to an integer for $N\rightarrow \infty$. Therefore, in the expression for the free energy in the large $N$ limit, one cannot separate the term which is singular near the critical point from the analytic background. Ref. \cite{Moshe:2003xn} gives the large-$N$ behavior of the free energy and of the amplitudes $R_c^+$ and $R_\xi^+$ for generic fixed dimension $d$; as explained, these results cannot be extended to $d=3$. 

Using the expression (\ref{a}), it is easy to obtain the large-$N$ limit of the following universal amplitude ratios
\begin{align}
U_0 = \displaystyle
\frac{\displaystyle -\frac{\pi^2}{32}\left( -\frac{24}{\pi^2} +3g''(0)+g'''(0) \right)}{\displaystyle 1+\frac{\pi^2}{32}\left( -\frac{24}{\pi^2} +3g''(0)+g'''(0) \right)},\\[1em]
R_c^+ = \displaystyle -\frac{\pi^2}{32}\left( -\frac{24}{\pi^2} +3g''(0)+g'''(0) \right).
\label{rc}
\end{align}

The leading behavior of $R_\xi^+$ is
\begin{equation}
R_\xi^+ = \left( \frac{R_c^+}{4 \pi} N \right)^{\frac{1}{3}},
\end{equation}
where $R_c^+$ is given by its large-$N$ limit (\ref{rc}).
The $1/N$ series for the coefficient $c_f$ is determined by the behavior of $f(x)$ and $g(x)$ near $x=-1$. We found the following expression
\begin{equation}
g(x)=\frac{12}{\pi^2} (x+1) \log (x+1) + C (x+1) + o(x+1),
\end{equation}
for $x \rightarrow -1$. The constant $C$ may be expressed as
\begin{equation}
\begin{array}{ll}
C= & \displaystyle - {\frac{4\,\left( 2 + \log (64) - 3\,\log (\pi ) \right) }{{{\pi }^2}}}\\[1em]
 & \displaystyle + \frac{2}{\pi^2}\,\int _{0}^{\infty} \left(
{\frac{6}{1 + k}} - {\frac{3\,{k^3}\,\pi }
         {\left( 1 + {k^2} \right) \,\left( 4 + {k^2} \right) \,\arctan ({\frac{k}{2}})}} \right)dk.
\end{array}
\end{equation}
Then the coefficient $c_f$ is (up to order $1/N$)
\begin{equation}
c_f=1+\frac{2}{N} \left( C-g(0) \right).
\end{equation}
The position of the crossover line can be determined by imposing $D'(y)=0$; solving this equation up to order $1/N$, we found
\begin{align}
x_{max} = & \displaystyle 5 + \frac{1}{N} 
\left( 64 - 30\,g'(5)\,{{\pi }^2} + 300\,g''(5)\,{{\pi }^2} + 5\,{{\pi }^2}\,g(5) \right),\\[1em]
y_{max} = & \displaystyle
{\frac{5}{{6^{{\frac{4}{5}}}}}} 
 + \frac{1}{N 15 \pi^2 6^{\frac{4}{5}}} 
\left( 320 - 150\,g'(5)\,{{\pi }^2} + 1500\,g''(5)\,{{\pi }^2} + 60\,{{\pi }^2}\,g(0)  \right. \\
\nonumber
&  \displaystyle  \left. + 15\,{{\pi }^2}\,g(5) +  16\,\log (216) \right), \\[1em] 
\nonumber
z_{max} = & \displaystyle
2\,{\sqrt{{\frac{3}{5}}}}+\frac{1}{N 5 \pi^2}\,{\sqrt{{\frac{3}{5}}}} 
 \left( -124 + 5\,g_0\,{{\pi }^2} + 30\,g'(5)\,{{\pi }^2} - 300\,g''(5)\,{{\pi }^2} \right.\\[1em]
 & \displaystyle \left. - 5\,{{\pi }^2}\,g(5) +40\, \log (5\pi/4) \right).
\end{align}
The universal amplitude ratios involving the crossover line are
\begin{align}
\nonumber
P_m = & \displaystyle {\frac{{\sqrt{5}}}{{6^{{\frac{2}{5}}}}}} 
+ \frac{1}{N 30\,{\sqrt{5}}\,{6^{{\frac{2}{5}}}}\,{{\pi }^2}}
 \left( 320 - 150\,g'(5)\,{{\pi }^2} + 1500\,g''(5)\,{{\pi }^2} + 60\,{{\pi }^2}\,g(0) \right.\\[1em]
& \displaystyle \left. + 15\,{{\pi }^2}\,g(5) + 
  528\,\log (6) - 600\,\log (5) \right),\\[1em]
\nonumber
P_c = & \displaystyle \frac{3125}{15552} - \frac{3125}{N 46656 \pi^2}
\left( -100 + 12\,{{\pi }^2} + 3\,g_0\,{{\pi }^2} + 30\,g'(5)\,{{\pi }^2} - 300\,g''(5)\,{{\pi }^2} \right. \\[1em]
& \displaystyle \left.  - 3\,{{\pi }^2}\,g(5) - 192\,\log (2) - 144\,\log (3) + 168\,\log (5) + 24\,\log (\pi ) \right),\\[1em]
\nonumber
R_p = & \displaystyle \frac{12}{5} + \frac{8}{N 25 \pi^2} 
\left( -76 + 15\,{{\pi }^2} + 15\,g'(5)\,{{\pi }^2} - 300\,g''(5)\,{{\pi }^2} + 180\,\log (5/6)  \right)
\end{align}
We report the results after calculating numerically the integrals:
\begin{align*}
r_6 & =  \displaystyle \frac{5}{6}+\frac{10.213}{N} & f^0_1 & =  \displaystyle 2-\frac{1.08789}{N} \\[1em]
r_8 & =  \displaystyle \frac{-67.314}{N} & f^0_2 & =  \displaystyle 1-\frac{1.89798}{N} \\[1em]
r_{10} & =  \displaystyle \frac{1406.83}{N} & f^0_3 & =  \displaystyle -\frac{0.642744}{N} \\[1em]
U_0 & =  \displaystyle 1.4674 & c_f & =  \displaystyle 1 + \frac{1.64657}{N} \\[1em]
R_\alpha & =  \displaystyle 0.4674 & P_m & =  \displaystyle 1.092 - \frac{0.313824}{N} \\[1em]
R_c^+ & =  \displaystyle 0.594715 & P_c & =  \displaystyle 0.200939+\frac{1.0391}{N} \\[1em]
R_4^+ & =  \displaystyle 12-\frac{54.0813}{N} & R_p & =  \displaystyle 2.4-\frac{3.70369}{N} \\[1em]
R_{\chi} & =  \displaystyle 1-\frac{1.91184}{N} + \frac{9.22}{N^2} & z_{max} & = \displaystyle 1.54919 - \frac{2.64874}{N} \\[1em]
R_{\xi}^+ & =  \displaystyle 0.361715 \, N^\frac{1}{3} & x_{max} & =  \displaystyle 5+ \frac{1.08647}{N} \\[1em]
F_0^\infty & =  \displaystyle \frac{1}{144}+\frac{0.0633049}{N} & y_{max} & =  \displaystyle 1.19247-\frac{0.515247}{N}.
\end{align*}
In the last equations we quoted from \cite{Oku:ht} the correction $1/N^2$ to the universal amplitude ratio $R_\chi$.

\end{document}